\documentclass{pasj00}

\begin{document}
\SetRunningHead{Goto et al.}{H$_3^+$ toward The Galactic Center III.}
\Received{2010/12/22}
\Accepted{2011/02/08}
\Published{2011/04/25}

\title{Absorption Line Survey of H$_3^+$ toward
  the~Galactic~Center~Sources III.  Extent of the Warm and
  Diffuse Clouds\footnotemark[\ast]}

\author{
  Miwa \textsc{Goto},\altaffilmark{1}
  Tomonori    \textsc{Usuda},\altaffilmark{2}
  Thomas R.   \textsc{Geballe},\altaffilmark{3}
  Nick \textsc{Indriolo},\altaffilmark{4}
  Benjamin J. \textsc{McCall},\altaffilmark{4}
  Thomas \textsc{Henning},\altaffilmark{1}
  and
  Takeshi     \textsc{Oka}\altaffilmark{5}}

\altaffiltext{1}{Max-Planck-Institut f\"ur Astronomie,
  K\"onigstuhl 17, D-69117 Heidelberg, Germany.}
\email{mgoto@mpia.de}

\altaffiltext{2}{Subaru Telescope, 650, North A'ohoku Place, Hilo,
                  HI 96720, USA.}

\altaffiltext{3}{Gemini Observatory,
                 670, North A'ohoku Place,  Hilo, HI 96720, USA.}

\altaffiltext{4}{Department of Astronomy and Department of
                 Chemistry, University of Illinois at Urbana-Champaign,
                 Urbana, IL 61801-3792, USA.}

\altaffiltext{5}{Department of Astronomy and Astrophysics,
                 Department of Chemistry, and Enrico Fermi Institute,
                 University of Chicago, Chicago, IL 60637, USA.}

\KeyWords{ISM: clouds --- ISM: lines and bands --- ISM: molecules
           --- Galaxy: center } 

\maketitle

\footnotetext[$\ast$]{Based on data collected at Subaru Telescope,
which is operated by the National Astronomical Observatory of Japan.}

\begin{abstract}

We present follow-up observations to those of \citet{geb10}, who found 
high column densities of H$_3^+$ $\sim$100~pc off of the Galactic center 
(GC) on the lines of sight to 2MASS~J17432173-2951430 (J1743) and 
2MASS~J17470898-2829561 (J1747). The wavelength coverages on these 
sightlines have been extended in order to observe two key transitions of 
H$_3^+$, $R$(3,3)$^l$ and $R$(2,2)$^l$, that constrain the temperatures 
and densities of the environments. The profiles of the H$_3^+$ 
$R$(3,3)$^l$ line, which is due only to gas in the GC, closely matches 
the differences between the H$_3^+$ $R$(1,1)$^l$ and CO line profiles, 
just as it does for previously studied sightlines in the GC. Absorption 
in the $R$(2,2)$^l$ line of H$_3^+$ is present in J1747 at velocities 
between $-$60 and $+$100~km~s$^{-1}$. This is the second clear detection 
of this line in the interstellar medium after GCIRS~3 in the Central 
Cluster. The temperature of the absorbing gas in this velocity range is 
350~K, significantly warmer than in the diffuse clouds in other parts of 
the Central Molecular Zone. This indicates that the absorbing gas is 
local to Sgr~B molecular cloud complex. The warm and diffuse gas 
revealed by \citet{oka05} apparently extends to $\sim$100~pc, but there 
is a hint that its temperature is somewhat lower in the line of sight to 
J1743 than elsewhere in the GC. The observation of H$_3^+$ toward J1747 
is compared with the recent Herschel observation of H$_2$O$^+$ toward 
Sgr B2 and their chemical relationship and remarkably similar velocity 
profiles are discussed.

\end{abstract}

\section{Introduction}

The study of the interstellar medium in the Galactic center using 
H$_3^+$ as an astrophysical probe was initiated by \citet{geb99}, who 
discovered that the Galactic center is the richest reservoir of H$_3^+$ 
in the Milky Way. The column density of H$_3^+$ toward the Galactic 
center is $>10^{15}$ cm$^{-2}$, an order of magnitude higher than in 
dense molecular clouds \citep{mcc99} and in diffuse clouds in the 
Galactic disk \citep{mcc02,ind07}. Studies of H$_3^+$ in the 
interstellar medium initially relied on the vibration-rotation 
transitions from the two lowest rotational levels, ($J$, $K$)=(1,1) and 
(1,0). The first interstellar H$_3^+$ line from a higher rotational 
level was discovered by \citet{got02} toward the bright infrared sources 
GCIRS~3 and GCS~3-2, near the Galactic center. The ($J$, $K$) = (3,3) 
level from which the observed $R$(3,3)$^l$ originates is 361~K above the 
lowest level, and is ``metastable'', that is, H$_3^+$ in this level does 
not decay by spontaneous emission \citep{pan86}.

\newpage

\citet{oka05} determined the physical state of the gas toward the bright 
star GCS~3-2 in the Galactic center based on a steady state calculation 
of thermalization by \citet{oka04}.  Their results are summarized as 
follows: (1) compared to diffuse clouds in the Galactic disk, the gas in 
the Galactic center producing absorption by H$_3^+$ has a similar 
density ($\leq$100~cm$^{-3}$), 
but a much higher temperature 
($\sim$250~K); (2) The path length of the absorbing cloud is at least 
30~pc, which means it occupies a significant portion of the volume of 
the Galactic center; (3) the cosmic ray ionization rate of H$_2$ must be 
at least 10$^{-15}$~s$^{-1}$, with the limiting value in effect if the 
absorbinng clouds' line of sight dimension is the full diameter of the 
Central Molecular Zone of the Galaxy (CMZ) \citep[200~pc]{mor96}.

Subsequently \citet{got08} found that the interpretation by 
\citet{oka05} is universal for sightlines toward 8 stars located in the 
Galactic center and extending from Sgr~A* to 30 pc east of Sgr~A*. 
Without exception all eight sightlines they observed show strong 
$R$(1,1)$^l$ and $R$(3,3)$^l$ absorptions with similar column densities 
of H$_3^+$ as toward GCS~3-2, demonstrating that the warm and diffuse 
gas is present over a wide region of the CMZ.

In order to explore a wider region of the CMZ, \citet{geb10} have 
undertaken a spectroscopic search, from 170~pc east to 170~pc west of 
Sgr~A*, for bright infrared stars in {\it Spitzer} GLIMPSE catalogue 
\citep {ram08} with clean infrared continua, that can serve as 
background sources for H$_3^+$ spectroscopy. They have reported 
strong H$_3^+$ $R$(1,1)$^l$ and CO v=2-0 absorption toward two stars 
thus selected, 2MASS~J17432173-2951430 and 2MASS~J17470898-2829561 
(hereafter J1743 and J1747), located 130~pc to the West and 80~pc to the 
East of Sgr~A*, respectively.  Their results strongly imply that the 
warm and diffuse environment revealed by \citet{oka05} extends to radii 
of $\sim$100~pc.

In the present study we have observed the key $R$(3,3)$^l$ and 
$R$(2,2)$^l$ absorptions toward J1743 and J1747 in order to better 
determine the properties of these two new sightlines. The distance to 
the Galactic center is assumed to be 7.6~kpc throughout this paper 
\citep{eis05,nis06}.


\begin{figure}
  \begin{center}
    \FigureFile(60mm,60mm){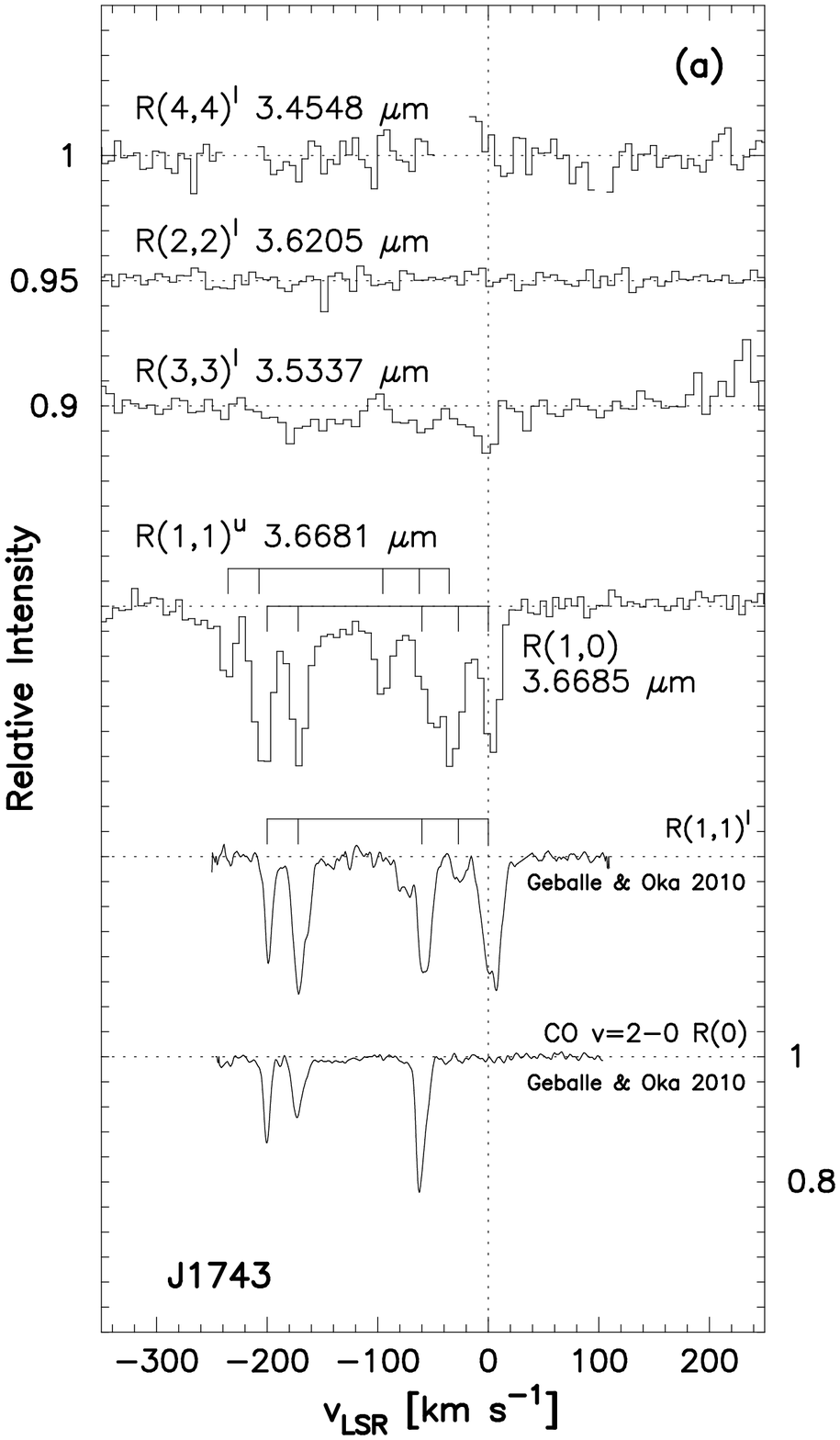}
    \FigureFile(60mm,60mm){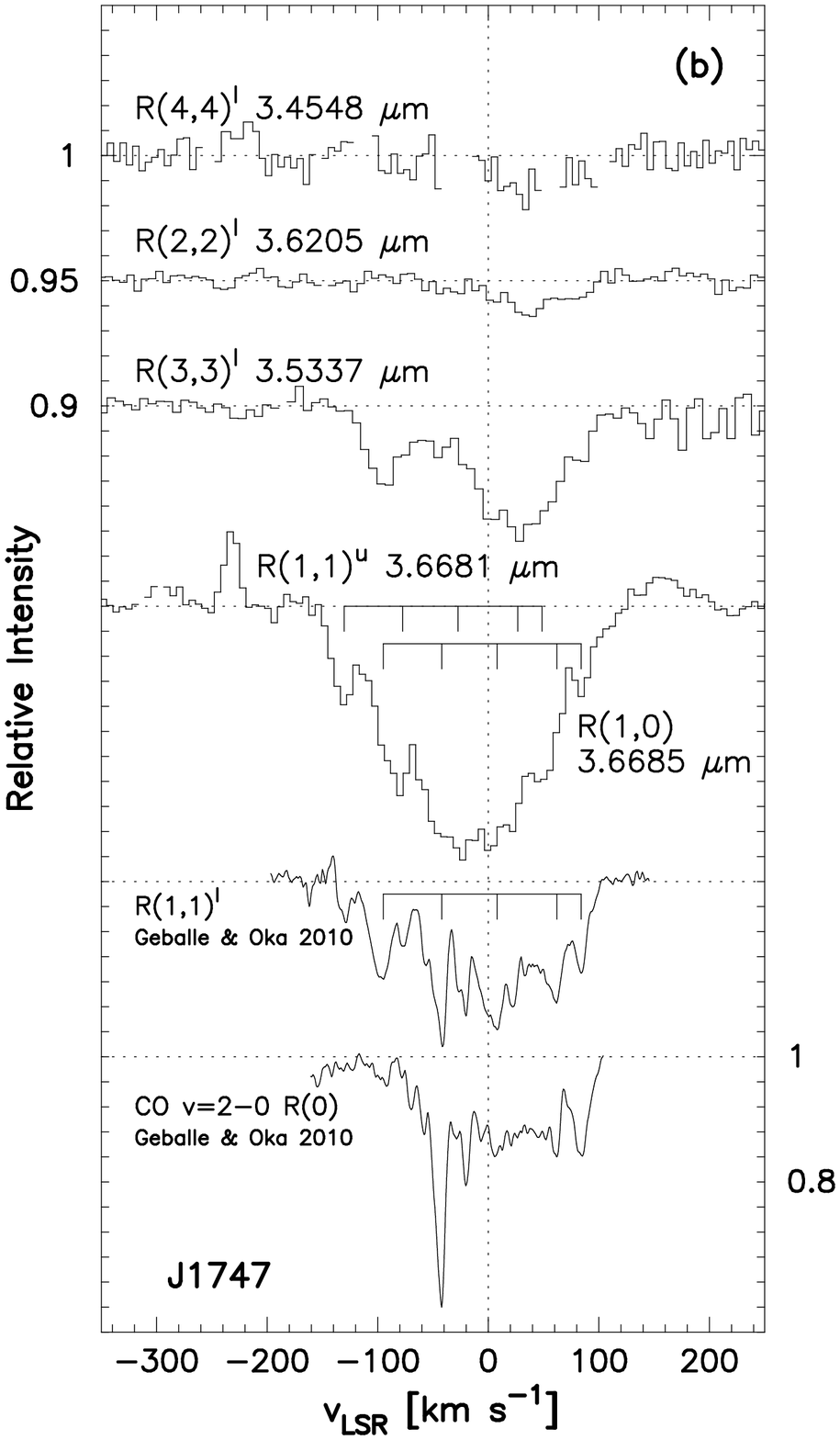}
  \end{center}
  \caption{Top (a):  the spectra of H$_3^+$ at $R$(4,4)$^l$,
    $R$(2,2)$^l$, $R$(1,0) [overlapped with $R$(1,1)$^u$] in the
    line of sight to 2MASS~J17432173-2951430. H$_3^+$
    $R$(1,1)$^l$ and CO $R$(0) lines are taken from
    \citet{geb10}. Botoom (b): same with the left, but for
    2MASS~J17470898-2829561. The velocities marked by the vertical lines are 0, $-$27, $-$60, $-$172, and $-$200~km~s$^{-1}$ for J1743, and 84, 62, 8, $-$42, and $-$95~km~s$^{-1}$ for J1747.}
\label{sp}
\end{figure}

\section{Observation and Data Reduction}

Spectra of J1743 and J1747 were obtained at the Subaru Telescope on UT 4 
May 2010 by the high-resolution infrared spectrograph IRCS 
\citep{kob00}. The resolving power used was $R$=20,000.  The angle 
settings of the echelle and cross-dispersing gratings were $-$9000 and 
2400, respectively, in order to cover H$_3^+$ $R$(4,4)$^l$ 
[3.4548~$\mu$m], $R$(2,2)$^l$ [3.6205~$\mu$m], $R$(1,0) [3.6685~$\mu$m], 
and $R$(1,1)$^u$ [3.6681~$\mu$m], and $-$200 and 4000, respectively, to 
cover $R$(3,3)$^l$ [3.5337~$\mu$m]. The data were obtained while nodding 
the telescope along the slit, whose dimensions were 
0\arcsec.14~$\times$6\arcsec.7, so as to be able to easily subtract the 
thermal sky background. The positions of the targets were continuously 
manually centered during the observation to maximize the throughput. The 
centering was performed with the camera of IRCS observing the $L^\prime$ 
band, the same wavelength as the spectroscopic observation, to minimize 
differential atmospheric refraction. An early-type spectroscopic 
standard star, HR~6378 (A2~V, $V=$2.43), was observed immediately before 
or after the object observation through similar airmass. Standard 
calibration data were obtained in the morning; these included 
spectroscopic flatfields with the same grating settings. The sky was 
clear and stable during the night, with the seeing measured as 
0\arcsec.59 in the $K$ band. The journal of observations is shown in 
Table~\ref{tb1}.

Data reduction was performed as described by \citet{got08}. The raw 
spectrograms were pair-subtracted, flat-fielded, and corrected for 
outlier pixels. One dimensional spectra of each of the Galactic center 
objects and of the standard star were extracted using IRAF\footnote{IRAF 
is distributed by the National Optical Astronomy Observatories, which 
are operated by the Association of Universities for Research in 
Astronomy, Inc., under cooperative agreement with the National Science 
Foundation.} aperture extraction package. The spectra of the Galactic 
center sources were further processed using custom-written IDL codes for 
correction for the telluric absorption lines, using the spectra of the 
standard star. Slight wavelength offsets, differences in airmass, 
fringes on the continuum, and saw-tooth features produced by different 
readout channels were simultaneously removed. Wavelength calibration was 
performed by maximizing the correlation with a model atmospheric 
transmission curve calculated using ATRAN \citep{lor92} and is better 
than 1~km~s$^{-1}$. Wavelengths were converted to radial velocities 
with respect to the local standard of the rest, using IRAF {\it rv} 
package. Normalized absorption spectra of H$_3^+$ are shown in 
Fig.~\ref{sp} for J1743 and J1747, together with the H$_3^+$ and CO 
spectra from \citet{geb10}.

\begin{figure}
  \begin{center}
    \FigureFile(90mm,60mm){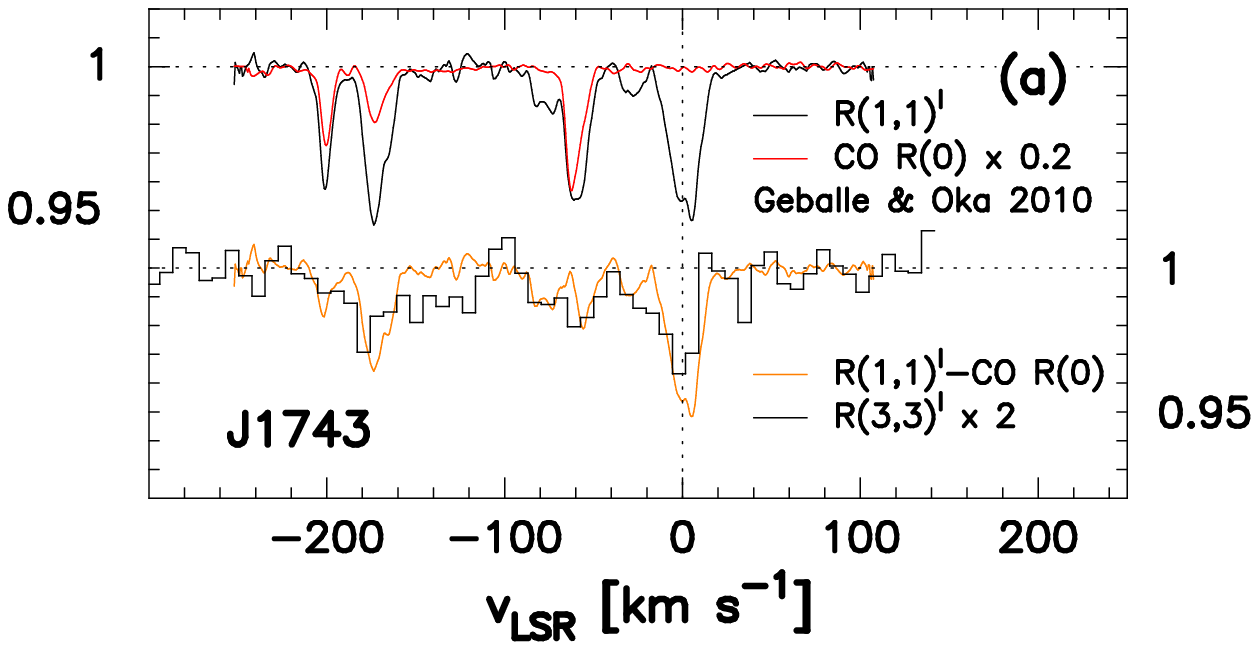}
    \FigureFile(90mm,60mm){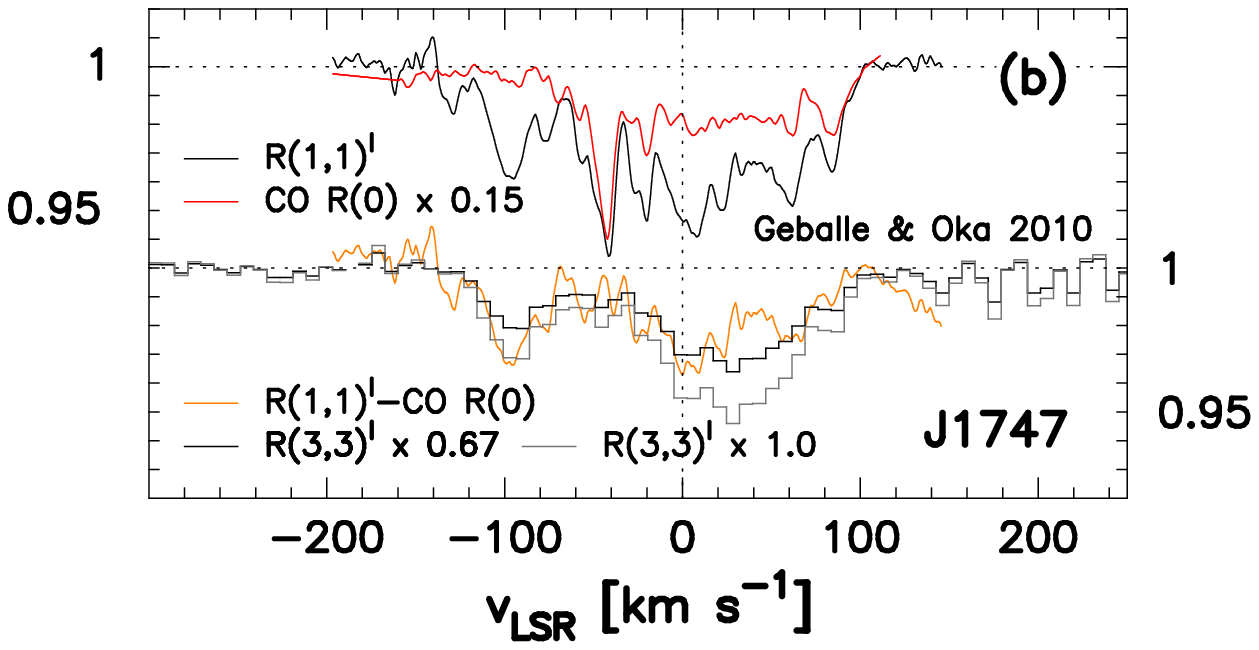}
  \end{center}
  \caption{Top (a): $R$(1,1)$^l$ and CO~$R$(0) spectra of
    2MASS~J17432173-2951430 taken from \citet{geb10}. CO~$R$(0)
    is scaled by the factor of 0.2 so that the sharp absorption
    lines at $-$60~km~s$^{-1}$ are roughly with the same depth.
    The difference of the two spectra is shown in orange in the
    lower trace to compare it with $R$(3,3)$^l$ spectrum
    obtained in this study. $R$(3,3)$^l$ spectrum is scaled by 2
    to match the differential spectra of $R$(1,1)$^l$ and
    CO~$R$(0). Bottom (b): same as the top, but for
    2MASS~J17470898-2829561. CO~$R$(0) is scaled by 0.15 so that
    the absorption line at $-$42~km~s$^{-1}$ is roughly the
    same depth with $R$(1,1)$^l$.  The differential
    spectrum of $R$(1,1)$^l$ and CO~$R$(0) is compared in the
    lower trace. $R$(3,3)$^l$ matches well with the differential
    spectrum at $>-$60~km~s$^{-1}$ after scaling by 0.67 (black
    trace), while the scaling factor is close to unity at
    $<-$60~km~s$^{-1}$ (gray trace).}

\label{df}
\end{figure}

\section{Results}

The spectra of H$_3^+$ $R$(1,1)$^l$ and the CO v=2-0 $R$(0) line in 
\citet{geb10} are considerably different from each other. Here we 
compare in detail the difference spectra of those two lines to the 
$R$(3,3)$^l$ line profiles observed here.

The upper traces of Fig.~\ref{df}a superimpose the velocity profiles of 
the H$_3^+$ $R$(1,1)$^l$ and CO~$R$(0) lines in J1743 taken from 
\citet{geb10}, but after rescaling CO~$R$(0) by a factor of 0.2 so that 
the common absorption components at $-$60~km~s$^{-1}$, which probably 
both arise solely in dense molecular gas in the foreground 3~kpc spiral 
arm as discussed by e.g., \citet{oka05}, are roughly the same depth. In 
J1743 the H$_3^+$ $R$(1,1)$^l$ line profile consists of six distinct 
absorptions, at 0, $-$27, $-$60, $-$75, $-$172, $-$200~km~s$^{-1}$, with 
minor substructure in the $-$172 and 0~km~s$^{-1}$ components 
\citep{geb10}.  The CO~v=2-0 line shows only three of these features: at 
$-$60, $-$172 and $-$200~km~s$^{-1}$. Perhaps most remarkably the CO 
profile is completely lacking the 0~km~s$^{-1}$ component, which is the 
strongest component in the H$_3^+$ $R$(1,1)$^l$ profile. The lower trace 
in orange represents the difference spectrum of $R$(1,1)$^l$ and the 
scaled CO $R$(0) overlaid with $R$(3,3)$^l$ from the present study. 
Despite being at lower spectral resolution than \citet{geb10}, the 
$R$(3,3)$^l$ profile qualitatively reproduces the difference spectrum, 
showing absorption components at 0, $-$75~km~s$^{-1}$ as well as the 
extra absorption at $-$172~ km~s$^{-1}$. The two cloud components with 
the high negative velocities of $-$200~km~s$^{-1}$ and 
$-$172~km~s$^{-1}$ are due to local dense clouds in Sgr~E as previously 
observed by \citet{lis92}.

The spectra of J1747 follow the same pattern. The upper traces of 
Fig.~\ref{df}b are the H$_3^+$ $R$(1,1)$^l$ and CO~2-0~$R$(1) lines in 
J1747 \citep{geb10}, after rescaling the CO line by a factor of 0.15, so 
that the common absorption components at $-$42~km~s$^{-1}$, which 
probably arise in dense cloud material in the foreground 3~kpc spiral 
arm, are the same depth. The absorption by CO~v=2-0 at negative 
velocities cuts off at $-$60~km~s$^{-1}$, unlike the absorption by 
H$_3^+$ $R$(1,1)$^l$, which extends to $-$150~km~s$^{-1}$. The 
absorptions by H$_3^+$ $R$(1,1)$^l$ and CO have similar extents at 
positive velocities but that of H$_3^+$ is much stronger than CO v=2-0 
after scaling. Again the $R$(3,3)$^l$ profile reproduces the extra 
absorption of $R$(1,1)$^l$ well, with two broad absorption components, 
from $-$120 to $-$60~km~s$^{-1}$ and from $-$60 to 100~km~s$^{-1}$. The 
rule that CO and H$_3^+$ $R$(3,3)$^l$ separately trace dense and diffuse 
gas in a line of sight, respectively, while $R$(1,1)$^l$ traces both of 
them \citep{oka05}, thus also applies to J1743 and J1747. This 
consistency lends support to the idea that the warm and diffuse clouds 
found on sightlines within 30~pc of the Galactic nucleus are present out 
to $\sim$100~pc.

Closer examination of the J1747 data reveals that the scaling factors of 
$R$(3,3)$^l$ needed to match $R$(1,1)$^l$ are slightly different at 
$<-$60~km~s$^{-1}$ and $>-$60~km~s$^{-1}$. The $R$(1,1)$^l$ $-$ 
CO~$R$(0) difference spectrum matches the $R$(3,3)$^l$ at 
$>-$60~km~s$^{-1}$ when scaling $R$(3,3)$^l$ by 0.67 (black trace in 
Fig.~\ref{df}b), while the scaling factor at $<-$60~km~s$^{-1}$ is close 
to unity (gray trace). This is most simply interpreted as a temperature 
effect. The $>-$60~km~s$^{-1}$ absorption is probably local to the Sgr~B 
star forming region \citep{geb10}, as is seen in the broad and strong 
absorption of CO~v=2-0. We infer that the absorbing gas in Sgr~B is 
warmer and thus produces a higher $n$(3,3)/$n$(1,1) ratio. The 
$<-$60~km~s$^{-1}$ absorption probably arises in the somewhat cooler 
diffuse clouds elsewhere in the CMZ.

\begin{figure}
  \begin{center}
    \FigureFile(80mm,60mm){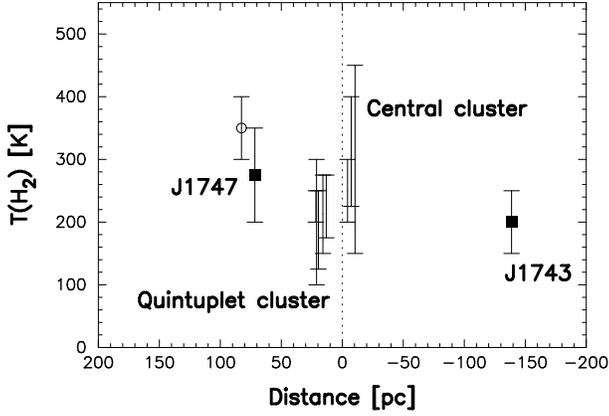}
  \end{center}
  \caption{Temperature distribution in the CMZ as a function of
    the distance from the Galactic nucleus.  The distance to the
    Galactic center is assumed to be 7.6~kpc.  The data for the
    Central cluster and the Quintuplet cluster are taken from
    \citet{got08}.  The filled squares are of the present study.
    The open circle is for the positive velocity component in
    2MASS~J17470898-2829561, which is likely locally associated
    to Sgr~B star forming region. }\label{l}
\end{figure}

The equivalent widths of the absorption lines in the warm and diffuse 
gas have been converted to the column densities in the v=0 (1,1), (2,2) 
and (3,3) levels in the same manner as \citet{geb99, got02}. The squares 
of transition dipole moments used for $R$(1,1)$^l$, $R$(2,2)$^l$, 
$R$(3,3)$^l$, and $R$(4,4)$^l$, 0.0141, 0.0177, 0.0191, and 
0.0198~D$^2$, respectively, are based on the Einstein A coefficients 
given in \citet{nea96}. The densities and the temperatures of the 
absorbing clouds were determined from the relative populations of 
H$_3^+$ in ($J$,$K$)=(1,1), (2,2) and (3,3), using the steady state 
model of \citet{oka04}. The detection of the $R$(2,2)$^l$ line toward 
J1747, only the second detection of this line, allows the density in 
this sightline to be determined, rather than an upper limit. The results 
are shown in Table~\ref{tb2}. Fig.~\ref{l} shows the temperature 
distribution in the CMZ as a function of the distance from the Galactic 
nucleus. The cloud temperatures observed so far in the CMZ are in the 
range 100--400~K. There is some possibility that temperatures are lower 
in the line of sight to J1743 than elsewhere.

\section{Discussion}

The region containing warm and diffuse gas is now 7 times more extensive 
in Galactic longitude than originally found by \citet{got08}. This 
environment remains exclusive to the CMZ of the Galactic center, as 
$R$(3,3)$^l$ has not been detected in the clouds in the Galactic disk. 
How far beyond a radius of 100~pc the outer boundary of the cloud is yet 
to be determined. Although the observations reported in this paper are 
only toward two stars, one 130~pc West and the other 80~pc East, 
additional observations toward several other stars between them (Oka et 
al., in preparation) yield comparable column densities of H$_3^+$ in 
warm and low density clouds, further suggesting that this environment 
exists throughout the CMZ with a large volume filling factor.

The recent detections of strong far infrared absorption by H$_2$O$^+$ 
toward Sgr~B2 \citep{sch10} using the HIFI instrument on the Herschel 
Space Observatory have given independent evidence for the existence of a 
large amount of diffuse gas in the CMZ. Because H$_2$O$^+$ is produced 
from OH$^+$ through the hydrogen abstraction reaction OH$^+$ + H$_2$ 
$\rightarrow$ H$_2$O$^+$ + H, and OH$^+$ is produced either from O$^+$ 
by the hydrogen abstraction reaction O$^+$ + H$_2$ $\rightarrow$ OH$^+$ 
+ H, or from O by the proton hop reaction O + H$_3^+$ $\rightarrow$ 
OH$^+$ + H$_2$, H$_2$O$^+$ is closely related chemically to H$_3^+$.  
H$_2$O$^+$ is rapidly destroyed by H$_2$ through hydrogen abstraction 
reaction H$_2$O$^+$ + H$_2$ $\rightarrow$ H$_3$O$^+$ + H, so it cannot 
be abundant in dense clouds.  \citet{ger10} and \citet{neu10} who 
observed strong absorption by both OH$^+$ and H$_2$O$^+$ toward W31C and 
W49N, respectively, have concluded from chemical analysis that these 
molecular ions are found in diffuse gas with a very low fraction of 
molecular hydrogen, $f($H$_2)$.  Unlike H$_3^+$ they cannot be used to 
determine temperatures directly because due to their large dipole 
moments, only the lowest levels are populated.  The ``spin" 
temperature determined from the observed ortho to para ratio of 
H$_2$O$^+$ is not straightforward to interpret in terms of a kinetic 
temperature. \citet{ger10} estimate $T\sim$ 100~K.

In addition to the chemical connection between H$_2$O$^+$ and H$_3^+$, 
there is some direct observational evidence that the two species are 
co-located.  The velocity profiles of ortho- and para-H$_2$O$^+$ 
reported by \citet{sch10} (see their Figs.~1 and 2) are remarkably 
similar to the H$_3^+$ velocity profile toward J1747 of \citet{geb10} 
reproduced in Figs.~1 and 2 of this paper.  Not only do both range from 
$\sim -$120~km~s$^{-1}$ to $\sim$90~km~s$^{-1}$, but also the velocities 
of individual peaks approximately match. The identical peaks at 
$-$48~km~s$^{-1}$, $-$26~km~s$^{-1}$, and 4~km~s$^{-1}$, which are due 
to the three foreground spiral arms, are not surprising. However, the 
absorption peaks near $-$100~km~s$^{-1}$ and 60~km~s$^{-1}$, which are 
due to gas within the GC, also match \citep{oka10}. Similar agreement is 
observed between the velocity profiles of $^{13}$CH$^+$ (E. Falgarone, 
private communication) and H$_3^+$. We regard these as strong 
indications that H$_2$O$^+$, CH$^+$ and H$_3^+$ are largely found 
together. This is somewhat surprising since H$_2$O$^+$ and CH$^+$ 
require low $f($H$_2)$ while H$_3^+$ favors high $f($H$_2)$.  The 
locations of Sgr~B2 and J1747 are separated in the plane of the sky by 
16~pc. The velocity correspondence may be due to the large extents of 
diffuse clouds within the Sgr B complex. 

\bigskip

We thank the staff of the Subaru Telescope and NAOJ for invaluable 
assistance in obtaining these data.  This research has made use of the 
SIMBAD database, operated at CDS, Strasbourg, France.  We also thank the 
referee, J.H. Black, for suggesting that we discuss relevant results 
from the Herschel Observation. T.R.G.'s research is supported by the 
Gemini Observatory, which is operated by the Association of Universities 
for Research in Astronomy, Inc., on behalf of the international Gemini 
partnership of Argentina, Australia, Brazil, Canada, Chile, the United 
Kingdom and the United States of America. B.J.M. and N.I. have been 
supported by NSF Grant PHY 08-55633. T.O. acknowledges NSF Grant AST 
08-49577.  We appreciate the hospitality of the local Hawaiian community 
that made possible the research presented here.




\onecolumn
\begin{table}
\footnotesize
  \caption{Journal of observations.}\label{tb1}

\begin{center}
\begin{tabular}{l ccc l ccc cc}
    \hline \hline
                             &   $l$    & $b$      & $L$     & Lines covered                                       & Exposure &\multicolumn{2}{c}{Grating$^\ast$}&\multicolumn{2}{c}{Standard}\\
                             & [\degree]& [\degree]&         &                                                     &      [s] & ECH   & XDP     &   Name          & Spe.         \\
\hline

2MASS~J17432173-2951430\dots & -1.046   &  -0.065  & 3.8     & $R$(4,4)$^l$, $R$(2,2)$^l$, $R$(1,1)$^u$, $R$(1,0)  &  960     &$-$9000&   2400  & HR~6378 & A2~V  \\
                             &          &          &         & $R$(3,3)$^l$                                   &  480     &$-$200 &   4000  & HR~6378 & A2~V  \\
                             &          &          &         &                                                     &          &       &         &         &       \\
2MASS~J17470898-2829561\dots &  0.548   & -0.059 &         & $R$(4,4)$^l$, $R$(2,2)$^l$, $R$(1,1)$^u$, $R$(1,0)  & 5400     &$-$9000&   2400  & HR~6378 & A2~V  \\
                             &          &          &         & $R$(3,3)$^l$                                   & 5400     &$-$200 &   4000  & HR~6378 & A2~V  \\

\hline

\end{tabular}
\end{center}
$^\ast$  ``ECH'' and ``XDP'' denote the motor settings for the angles of echelle and cross-dispersing gratings in the instrumental unit. \\

\end{table}

\begin{table}
  \caption{Column densities of H$_3^+$ toward 2MASS~1747 and 2MASS~1743.\label{tb2}}
\tiny

\begin{center}
\begin{tabular}{l c cccc cc cc}
    \hline \hline  
                             & $\Delta v^a$     & $N$(1,1)             & $N$(3,3)             & $N$(2,2)$^d$         & $N$(4,4)$^d$         & $n$(3,3)/$n$(1,1) & $n$(3,3)/$n$(2,2) & $n(\rm{H_2})$ & $T(\rm{H_2})$ \\
                             & [km~s$^{-1}$]    & [10$^{14}$cm$^{-2}$] & [10$^{14}$cm$^{-2}$] & [10$^{14}$cm$^{-2}$] & [10$^{14}$cm$^{-2}$] &                   &                   & [cm$^{-3}$]   & [K]           \\ 
\hline

2MASS~J17470898-2829561\dots & $-$150,$-$60$^b$ & 6.0$\pm$0.7          & 4.3$\pm$1.3          & $<$0.8               & $<$1.8               & 0.7$\pm$0.2       & $>$5.4            & $<$50         & 200--350      \\  
                             & $-$60,$+$100$^c$ & 16.4$\pm$1.2         & 18.2$\pm$2.2         & 5.8$\pm$1.5          & $<$3.3               & 1.1$\pm$0.2       & 3.1$\pm$0.9       & 100--150      & 300--400      \\  
                             &                  &                      &                      &                      &                      &                   &                   &               &               \\  

2MASS~J17432173-2951430\dots & $-$220,$+$20     & 12.8$\pm$1.9         & 5.7$\pm$.2.2         & $<$2.1               & $<$5.0               & 0.45$\pm$0.2      & $>$2.9            & $<$50         & 150--250      \\  
                             &                  &                      &                      &                      &                      &                   &                   &               &               \\  

\hline

\end{tabular}
\end{center}
$^a$  The interval that the column densities are calculated.
$^b$  The CMZ component.
$^c$  Sgr~B component. 
$^d$  The upper limits are for 1 $\sigma$.

\end{table}

\end{document}